\documentclass[10pt,twocolumn,letterpaper]{article}

\usepackage[pagenumbers]{cvpr} 

\usepackage{graphicx}
\usepackage{amsmath}
\usepackage{amssymb}
\usepackage{booktabs}
\usepackage{multirow}

\usepackage[pagebackref,breaklinks,colorlinks]{hyperref}
\usepackage{algorithm, algorithmic}

\usepackage[capitalize]{cleveref}
\crefname{section}{Sec.}{Secs.}
\Crefname{section}{Section}{Sections}
\Crefname{table}{Table}{Tables}
\crefname{table}{Tab.}{Tabs.}

\begin{document}

\title{CIT-EmotionNet: CNN Interactive Transformer Network for EEG Emotion Recognition}

\author{%
  Wei Lu$^{1,2}$ \quad Hua Ma$^{2}$ \quad  Tien-Ping Tan$^{1,}$* \\
  $^{1}$ Universiti Sains Malaysia \quad $^{2}$ Zhengzhou Railway Vocational and Technical College \quad 
  \\
}
\maketitle

\begin{abstract}
Emotion recognition using Electroencephalogram (EEG) signals has emerged as a significant research challenge in affective computing and intelligent interaction. However, effectively combining global and local features of EEG signals to improve performance in emotion recognition is still a difficult task. In this study, we propose a novel CNN Interactive Transformer Network for EEG Emotion Recognition, known as CIT-EmotionNet, which efficiently integrates global and local features of EEG signals. Initially, we convert raw EEG signals into spatial-frequency representations, which serve as inputs. Then, we integrate Convolutional Neural Network (CNN) and Transformer within a single framework in a parallel manner. Finally, we design a CNN interactive Transformer module, which facilitates the interaction and fusion of local and global features, thereby enhancing the model's ability to extract both types of features from EEG spatial-frequency representations. The proposed CIT-EmotionNet outperforms state-of-the-art methods, achieving an average recognition accuracy of 98.57\% and 92.09\% on two publicly available datasets, SEED and SEED-IV, respectively.
\end{abstract}

\section{Introduction}
Emotion recognition \cite{jia2021hetemotionnet, tan2020fusionsense, cimtay2020cross, doma2020comparative} has become a significant research task in the field of affective computing due to its potential applications in various areas, including affective brain-computer interfaces, the diagnosis of affective disorders, emotion detection in patients with consciousness disorders, emotion detection of drivers, mental workload estimation, and cognitive neuroscience. Emotion is a mental and physiological state that arises from various sensory and cognitive inputs, significantly influencing human behavior in daily life \cite{jia2021hetemotionnet}. Emotion is a response to both internal and external stimuli. Physiological signals, such as Electrocardiogram (ECG), Electromyography (EMG), and EEG, correspond to the physiological responses caused by emotions and are considered more reliable indicators of emotional expression than non-physiological signals, such as speech, posture, and facial expression, which can be masked by humans \cite{tan2020fusionsense,cimtay2020cross}. Among physiological signals, EEG signals provide high temporal resolution and rich information that can reveal subtle changes in emotions, making them more suitable for emotion recognition than other physiological signals \cite{atkinson2016improving}. EEG-based emotion recognition methods are more accurate and objective, as some studies have verified the relationship between EEG signals and emotions \cite{xing2019sae+}. However, effectively combining global and local features of EEG signals to achieve better performance in emotion recognition remains a challenge.

The method based on convolutional neural network (CNN)~\cite{kwon2018electroencephalography, shen2023triplet, salama2018eeg, hpgn} is well-known for its efficient feature extraction and powerful feature representation ability. Therefore, an increasing number of researchers have begun exploring the use of CNN models for emotion recognition from electroencephalography (EEG) signals. For example, Kwon et al. introduced a CNN model for EEG feature extraction \cite{kwon2018electroencephalography}. Yang et al. proposed a multi-column CNN-based model for emotion recognition from EEG signals \cite{yang2019multi}. Chen et al. presented a CNN-based method for learning and classifying EEG emotion features \cite{chen2019accurate}. However, these methods are limited in capturing temporal information and global features \cite{li2022eeg}.

With the rapid development of Transformer~\cite{liu2022spatial}, it has been successfully applied in many fields. Shen et al \cite{git}  proposed a structure where graphs and transformers interact constantly, enabling close collaboration between global and local features. In the area of emotion recognition, existing work typically uses classic backbone networks for feature extraction. For example, Liu et al~\cite{liu2022spatial} proposed an EEG emotion Transformer (EeT) model that directly learns spatial-spectral features from EEG signal sequences, making the traditional Transformer model suitable for EEG signals \cite{liu2022spatial}. However, these models are not adept at handling local features. To enhance the accuracy of EEG emotion recognition, researchers are investigating effective ways to combine the strengths of both CNN and Transformer models.

In order to address the above challenging task, we propose a CNN interactive Transformer network for EEG emotion recognition named CIT-EmotionNet. First, we divided the original EEG signal into several segments of 4 seconds each. For each segment, we extracted the PSD features of the $\delta$, $\theta$, $\alpha$, $\beta$, and $\gamma$ frequency bands, and mapped the features in space according to the location of the electrodes to generate the EEG feature representation. In order to extract both local and global features of the EEG feature representation, we sent the EEG feature representation to the CNN branch and the Transformer branch, respectively. Afterward, in order to better fuse and interact with the local features extracted by the CNN branch and the global features extracted by the Transformer branch, we designed a CNN interactive Transformer module, named the CIT module. The classifier is composed of fully connected layers and softmax layers, which are used to predict the emotional label. Finally, we conducted a series of comparative experiments and ablation experiments using CIT-EmotionNet, which not only proved that CIT-EmotionNet outperformed all state-of-the-art models, but also studied the contribution of key components in CIT-EmotionNet to recognition performance.

The main contributions of this paper can be summarized as follows:

\begin{itemize}
\item	To extract both local and global features of EEG signals, we proposed a CNN interactive Transformer Network for EEG Emotion Recognition, named CIT-EmotionNet. It effectively utilizes the advantages of CNN in extracting local features and Transformer in extracting global features and unifies them in a single framework to form a powerful model.
\item	To better fuse and interact with local and global features, we have developed a CIT module that enables the interaction and fusion of both types of features. This has improved the model's ability to extract both local and global features from EEG feature representations.
\item	The proposed CIT-EmotionNet model achieves an average recognition accuracy of 98.57\% and 92.09\% on the SEED and SEED-IV datasets, respectively, outperforming all state-of-the-art models. In addition, we conducted a series of ablation experiments to investigate the contributions of the key components of CIT-EmotionNet to the recognition performance.
\end{itemize}

\section{METHODOLOGY}
In this section, the proposed network is presented in six parts: preliminaries, model overview, EEG feature representation, CNN branch, Transformer branch, and CIT module.

\subsection{Overview}
In this paper, we define $S = ({S_1},{S_2},...,{S_B}) \in {\mathbb{R}^{{N_e} \times B}}$ as the frequency features containing $B$  frequency bands extracted from EEG signals, where ${N_e}$ is the number of electrodes. We construct the spatial-spectral representation $S_{}^M = (S_1^M,S_2^M,...,S_B^M) \in {\mathbb{R}^{H \times W \times B}}$ , where $H$ and $W$ denote the height and width of the frequency map. $\left\{ {\delta ,\theta ,\alpha ,\beta ,\gamma } \right\}$ is selected as the frequency band set.

This section presents an overview of the proposed CIT-EmotionNet model structure. The raw EEG signals are first converted into EEG feature representations. Subsequently, these EEG feature representations are fed into CIT-EmotionNet for EEG emotion recognition. CIT-EmotionNet is primarily composed of three parts: CNN branch, Transformer branch, and CIT module. The CNN branch consists of five stages, namely stage 0, stage 1, stage 2, stage 3, and stage 4. Each stage of the CNN branch from stage 1 to stage 4 contains two ResNet basic blocks, which respectively output local features C1, C2, C3, and C4. Similarly, the Transformer branch also consists of five stages, namely stage 0, stage 1, stage 2, stage 3, and stage 4. Each stage of the Transformer branch from stage 1 to stage 4 contains three Encoder blocks, which respectively output global features T1, T2, T3, and T4. The local features C1, C2, and C3, as well as the global features T1, T2, and T3, are input into the CIT module of stages 1, 2, and 3, where they are interactively fused. The CIT module comprises two blocks: the global to local block and the local to global block. The output of the CIT module for each stage is the input for the next stage of both the CNN branch and the Transformer branch. Specifically, the output of the global to local block corresponds to the input for the next stage of the CNN branch, while the output of the local to global block corresponds to the input for the next stage of the Transformer branch. Finally, the local features C4 from stage 4 of the CNN branch and the global features T4 from stage 4 of the Transformer branch are merged through a Concat module, which are then fed into a fully connected layer for emotion classification recognition.

\subsection{EEG Feature Representation}
We divide the original EEG signals into non-overlapping periods lasting for 4 seconds, and each segment is assigned the same label as the original EEG signals.

To construct the EEG feature representation, the temporal-frequency feature extraction method is used to extract the Power Spectral Density (PSD) features of five frequency bands $\left\{ {\delta ,\theta ,\alpha ,\beta ,\gamma } \right\}$ of all EEG channels from the EEG signal samples in the EEG segments with a length of 4 seconds. We define $S = ({S_1},{S_2},...,{S_B}) \in {\mathbb{R}^{{N_e} \times B}}$ as a frequency feature containing a frequency band extracted from the PSD feature, in which the frequency band is $B \in \left\{ {\delta ,\theta ,\alpha ,\beta ,\gamma } \right\}$ , the electrode is ${N_e} \in \{ FP1,FPZ,...,CB2\}$ , and $X_b^B = (x_b^1,x_b^2,...,x_b^N) \in {\mathbb{R}^N},(b \in \{ 1,2,...,B\} )$ represent the collection of EEG signals from all ${N_e}$ electrodes on the frequency band $B$. Then, the selected data are mapped to a frequency domain brain electrode location matrix $S_b^M \in {\mathbb{R}^{H \times W}},(b \in \{ 1,2,...,B\} )$  according to the electrode location on the brain. Finally, the frequency-domain brain electrode position matrices from different frequencies are superimposed to form the spatial-frequency feature representation of EEG signals, that is, the construction of the EEG feature representation $S_{}^M = (S_1^M,S_2^M,...,S_B^M) \in {\mathbb{R}^{H \times W \times B}}$ is finished.

\subsection{CNN Branch}
The Residual Network (ResNet) was proposed to address the problem of degradation in deep CNN models. ResNet utilizes residual connections to link different convolutional layers, thereby enabling the propagation of shallow feature information to the deeper layers. As a result of these advantages of ResNet, we have chosen it as the local feature extraction model for the CNN branch.

The input of ResNet is the spatial-spectral feature representation $S_{}^M = (S_1^M,S_2^M,...,S_B^M) \in {\mathbb{R}^{H \times W \times B}}$. The spatial-spectral feature representation first goes through Stage 0, which consists of a $7 \times 7$ convolutional layer, a max pooling operation, and Batch Normalization. The purpose of this stage is to perform initial convolutional feature extraction and downsampling of the EEG feature representation, reducing the size of the feature map while increasing the level of abstraction. Stage 0 preprocesses the input EEG feature representation through convolution, pooling, and Batch Normalization operations, transforming it into a smaller and more abstract feature representation, providing better input features for subsequent convolutional blocks. Specifically, the input of Stage 0 is the spatial-spectral feature representation $S_{}^M = (S_1^M,S_2^M,...,S_B^M) \in {\mathbb{R}^{H \times W \times B}}$, where the shape of the spatial-spectral feature representation is $H \times W \times C$, with $H$ representing the height, $W$ representing the width, and $C$ representing the number of channels. Since the number of frequency bands is $5$, $C=5$. However, this does not meet the input requirements of the original ResNet model since the first convolutional layer in the original ResNet model requires $3$ input channels. If we directly use the original model to process data with $5$ input channels, channel conversion or padding operations are required, which can be tedious. Therefore, we replaced the first half of the ResNet model with a new convolutional layer that has $5$ input channels, $64$ output channels, a kernel size of $7 \times 7$, a stride of $2$, a padding of $4$, and no bias. After passing through the new convolutional layer, the size and number of the output feature maps have changed. Specifically, after passing through the new convolutional layer, the shape of the output feature map is $\frac{H+8}{2} \times \frac{W+8}{2} \times 64$, where $+8$ is due to the padding of $4$. This feature map will be used as the input for the subsequent Stage 1. The equations for Stage 0 of ResNet are shown in equations Equation~(\ref{eq1}).

\begin{equation}\label{eq1}
\resizebox{.9\hsize}{!}{${C_{0}} = \mathrm{MaxPool}(\mathrm{ReLU}(\mathrm{BN}(\mathrm{Conv_{7\times7}}({S_{}^M}))))$}
\end{equation}

 Where $S_{}^M$ is the input of stage 0 in the CNN branch, $C0$ is the output of Stage 0 in the CNN branch. $\mathrm{Conv _{7\times7}}(\cdot)$ represents the convolutional layer operation with an output channel of $64$, kernel size of $7 \times 7$, the stride of $2$, and padding of $4$. $\mathrm{BN}(\cdot)$ represents the batch normalization layer operation, which performs batch normalization on the output of the convolutional layer. $\mathrm{ReLU}(\cdot)$ represents the ReLU activation function, which applies the ReLU activation function to the output of the batch normalization layer. $\mathrm{MaxPool}(\cdot)$ represents the max pooling layer operation, which performs max pooling using a $3 \times 3$ pooling kernel, a stride of $2$, and padding of $1$.

The features output from Stage 0 are processed through Stage 1, Stage 2, Stage 3, and Stage 4 to generate the C1, C2, C3, and C4 features, respectively. After passing through Stage 1, the shape of the output feature C1 remains the same as the input feature, which is $\frac{H+8}{2} \times \frac{W+8}{2} \times 64$. After passing through Stage 2, the shape of the output feature C2 is $\frac{H+16}{4} \times \frac{W+16}{4} \times 128$. After passing through Stage 3, the shape of the output feature C3 is $\frac{H+32}{8} \times \frac{W+32}{8} \times 256$. After passing through Stage 4, the shape of the output feature C4 is $\frac{H+64}{16} \times \frac{W+64}{16} \times 512$. Each of Stage 1, Stage 2, Stage 3, and Stage 4 consists of 2 BasicBlocks. In BasicBlock, the input feature is added to the main branch output feature via a shortcut connection before being passed through a ReLU activation function. The equation of the main branch, as shown in Equation~(\ref{eq2}).

\begin{equation}\label{eq2}
\resizebox{.8\hsize}{!}{$X_{main} = \mathrm{BN}(\mathrm{Conv_{3\times3}}(\mathrm{ReLU}(\mathrm{BN}(\mathrm{Conv_{3\times3}}({X_{BasicIN}})))))$},
\end{equation}

Where $X_{BasicIN}$ is the input of BasicBlock, $X_{main}$ is the output of the main branch in the BasicBlock. $\mathrm{Conv_{3\times3}}(\cdot)$ represents the convolutional layer operation with a kernel size of $3 \times 3$, the stride of $1$, and padding of $1$. $\mathrm{BN}(\cdot)$ represents the batch normalization layer operation, which performs batch normalization on the output of the convolutional layer. $\mathrm{ReLU}(\cdot)$ represents the ReLU activation function, which applies the ReLU activation function to the output of the batch normalization layer. 

The shortcut connection allows the gradient to flow directly through the network, bypassing the convolutional layers in the main branch, which helps to prevent the vanishing gradient problem. The equation of the shortcut connection, as shown in Equation~(\ref{eq3}).
\begin{equation}\label{eq3}
X_{shortcut} = \mathrm{BN}(\mathrm{Conv_{1\times1}}(\mathrm{Conv_{3\times3}}({X_{BasicIN}}))),
\end{equation}
Where $X_{BasicIN}$ is the input of BasicBlock, $X_{shortcut}$ is the output of the shortcut connection in the BasicBlock. $\mathrm{Conv_{3\times3}}(\cdot)$ represents the convolutional layer operation with a kernel size of $3 \times 3$, the stride of $1$, and padding of $1$. $\mathrm{Conv_{1\times1}}(\cdot)$ represents the convolutional layer operation with a kernel size of $1 \times 1$. $\mathrm{BN}(\cdot)$ represents the batch normalization layer operation. $\mathrm{ReLU}(\cdot)$ represents the ReLU activation function.

The addition of the input feature to the main branch output feature allows the network to learn residual mappings, which can be easier to optimize during training. The equation of the addition, as shown in Equation~(\ref{eq4}).
\begin{equation}\label{eq4}
X_{BasicOUT} =\mathrm{ReLU}(X_{main}+X_{shortcut}),
\end{equation}
Where $X_{main}$ is the output of the main branch, $X_{shortcut}$ is the output of the shortcut connection, and $X_{BasicOUT}$ is the output of the BasicBlock.  $\mathrm{ReLU}(\cdot)$ represents the ReLU activation function.

\subsection{Transformer Branch}
Recently, Vision Transformer (ViT) has shown great potential in various computer vision applications, such as image classification and segmentation. Therefore, we chose ViT to extract global features for the Transformer branch.

The input of ViT is the spatial-spectral feature representation $S_{}^M = (S_1^M,S_2^M,...,S_B^M) \in {\mathbb{R}^{H \times W \times B}}$. The shape of $S_{}^M$ is $32  \times 32  \times 5$. $S_{}^M$ is first processed by Stage 0 and segmented into small blocks, each called a patch. Then, each patch is compressed into a one-dimensional vector, which serves as the input for subsequent self-attention calculations. We set $ patch\_size=32 $ and $ hidden\_dim=256 $. Therefore, $S_{}^M$ with a shape of $32 \times 32 \times 5$ is divided into 5 patches of size $(32\times 32)$ each, with a shape of $32 \times 32 \times 1$. After Stage 0, each patch is compressed into a vector of size $256$, resulting in an output shape of Stage 0 of $5 \times 256$.

The features output from Stage 0 are processed through Stage 1, Stage 2, Stage 3, and Stage 4 to generate the T1, T2, T3, and T4 features, respectively. ViT's Stage 1, Stage 2, Stage 3, and Stage 4 each contain 3 EncoderBlocks. The EncoderBlock in ViT is designed to establish global context relationships between different positions of the input feature map so that the model can better understand the input feature map. An EncoderBlock comprises of three primary components: Multi-Head Attention, Multi-Layer Perceptron (MLP) Block, and Layer Normalization. Below is a detailed description of each of these components.

Multi-Head Attention is utilized to perform self-attention operations in different representation subspaces to capture global relationships. Specifically, the Multi-Head Attention first maps the input vector $X$ into three sequences of vectors, namely, $Q\in\mathbb{R}^{n\times d_q}$, $K\in\mathbb{R}^{n\times d_k}$, and $V\in\mathbb{R}^{n\times d_v}$, by three different linear transformations, namely, Query, Key, and Value matrices. Here, $d_q$, $d_k$, and $d_v$ are the dimensions of the Query, Key, and Value vectors, respectively. Then, $Q$, $K$, and $V$ are transformed by linear mappings of dimension $d_h$ to obtain multiple heads $H$, as shown in Equation (\ref{eq5}).

\begin{equation}\label{eq5}
hea{d_i} = {\mathrm{Attention}}(QW_i^Q,KW_i^K,VW_i^V),
\end{equation}
Wher, $W_i^Q\in\mathbb{R}^{d_q\times d_h}$, $W_i^K\in\mathbb{R}^{d_k\times d_h}$, and $W_i^V\in\mathbb{R}^{d_v\times d_h}$ are parameter matrices of the $i$-th head, and $\mathrm{Attention}(\cdot)$ denotes the dot-product attention mechanism. 

Dot-Product Attention is used to compute the similarity between Q and K and weight V by this similarity, as shown in Equation (\ref{eq6}).

\begin{equation}\label{eq6}
\mathrm{Attention}(Q,K,V) = {\mathrm{Softmax}}\left( {\frac{{Q{K^T}}}{{\sqrt {{d_k}} }}} \right)V,
\end{equation}
Where, $Q$, $K$, and $V$ represent the query, key, and value vectors, respectively, and $d_k$ is the dimension of the key vector. The function $\mathrm{softmax}(\cdot)$ is used to normalize the similarity between the query and key vectors, obtaining the attention weights, which are then used to obtain the weighted sum of the value vectors, resulting in the final output.

To enable the model to learn features in different subspaces and increase its expressive power, the Multi-Head Attention concatenates the output vectors of these attention heads together, and then maps them to a new space through a linear transformation layer, as shown in Equation (\ref{eq7}).

\begin{equation}\label{eq7}
\resizebox{.8\hsize}{!}{${\text{MultiHead}}(Q,K,V) = {\text{Concat}}(hea{d_1}, \cdots ,hea{d_h}){W^O}$},
\end{equation}

Where, $head_i$ represents the output of the $i$-th attention head, $h$ denotes the number of heads, $\text{Concat}(\cdot)$ is the operation that concatenates the output of each head, and $W^O$ is the linear transformation matrix for the final output. 

The MLP Block refers to a multi-layer perceptron module, also known as a fully connected layer module, usually consisting of two linear layers and a GELU activation function. Specifically, the input to each MLP Block is a tensor processed by the Multi-Head Attention module, with a shape of $N \times L \times D$, where $N$ represents the number of input sequences, $L$ represents the length of the sequence, and $D$ represents the vector dimension at each position. Then, the MLP Block transforms each vector at each position through two fully connected layers and uses GELU as the activation function. The MLP Block can not only improve the expressive power of the model but also ensure that information is not lost during processing. The MLP Block is shown in Equation (\ref{eq8}).

\begin{equation}\label{eq8}
\resizebox{.8\hsize}{!}{$\mathrm{MLP}(x) = \mathrm{Dropout}(\mathrm{Dropout}(\mathrm{GELU}(x{W_1} + {b_1})){W_2} + {b_2})$},
\end{equation}

Where, $x$ is the input vector, $W_1$ and $W_2$ are the weights of the two fully connected layers, $b_1$ and $b_2$ are the bias terms, $\mathrm{GELU}(\cdot)$ represents the GELU activation function, and $\mathrm{Dropout}(\cdot)$ represents the Dropout layer, which is used to reduce overfitting.

Layer Normalization is used to normalize inputs to improve the stability and training effectiveness in MLP and Multi-Head Attention modules. Specifically, Layer Normalization normalizes inputs by subtracting the mean and dividing by the standard deviation of the features. This approach effectively combats the problem of vanishing gradients, without reducing network expressivity, thereby enhancing network stability. The formula of Layer Normalization is shown in Equation (\ref{eq9}).

\begin{equation}\label{eq9}
\mathrm{LayerNorm(x)} = \gamma \frac{{x - \mu }}{{\sqrt {{\sigma ^2} + \varepsilon } }} + \beta ,
\end{equation}
Where, $x$ represents the input features, $\gamma$, and $\beta$ are learnable scaling and shifting parameters, $\mu$ and $\sigma$ are the mean and standard deviation of the input vector $x$, and $\epsilon$ is a very small value used to avoid the denominator being zero.

\subsection{CIT Module}
The local features extracted by the CNN branch can provide more fine-grained information, while the global features extracted by the Transformer branch can provide more comprehensive contextual information. To better integrate local and global features and thus improve the accuracy and robustness of the model, we propose a CNN interactive Transformer module, which is called the CIT module. The CIT module enables the interaction and fusion of local and global features, thus enhancing the ability of the model to extract both global and local features from EEG spatial-spectral feature representation, leading to improved performance.

Every CIT module consists of two blocks: the Local to Global block, termed L2G block, and the Global to Local block, termed G2L block. The CNN feature maps (Stage $k$, where $k=1,2,3$) serve as inputs to the L2G block, which outputs ViT feature maps (Stage $k+1$, where $k=1,2,3$). Similarly, the ViT feature maps (Stage $k$, where $k=1,2,3$) serve as inputs to the G2L block, which outputs CNN feature maps (Stage $k+1$, where $k=1,2,3$). The CIT module can make the local features of the CNN feature map and the global features of the ViT feature map form a coupling state. In other words, global features and local features can interact with each other in the entire feature learning process. The following is a detailed introduction to the L2G block and the G2L block. 

In order to improve the ability of the model to extract features from different regions of EEG spatial-spectral feature representation and thereby improve the classification performance, we propose a CNN interactive Transformer module, termed the CIT module. The CIT module enables the interaction and fusion of local and global features, thus enhancing the ability of the model to extract both global and local features from EEG spatial-spectral feature representation, leading to improved performance. The overall framework of the CIT module is shown in Figure 5. Every CIT module consists of two blocks: the Local to Global block, termed L2G, and the Global to Local block, termed G2L. The CNN feature maps (Stage $k$, where $k=1,2,3$) serve as inputs to the L2G block, which outputs ViT feature maps (Stage $k+1$, where $k=1,2,3$). Similarly, the ViT feature maps (Stage $k$, where $k=1,2,3$) serve as inputs to the G2L block, which outputs CNN feature maps (Stage $k+1$, where $k=1,2,3$). The CIT module can make the local features of CNN feature map and the global features of ViT feature map form a coupling state. In other words, global features and local features can interact with each other in the entire feature learning process. The following is a detailed introduction to the L2G block and the G2L block.

\subsubsection{L2G Block}

The L2G block is responsible for transforming the input CNN feature maps into ViT feature maps. This block consists of Mean, Linear, Expand, and Concat, and the processing of the L2G block is shown in the following equation.

\begin{equation}\label{eq10}
F_G^{k + 1} = \mathrm{L2G}(F_L^k), k \in \left\{ {1,2,3} \right\},
\end{equation}
where, $F_G^{k+1}$ represents the global feature at stage $k+1$ and serves as the output to the L2G block, $F_L^k$ represents the local feature at stage $k$ and serves as the input of the L2G block, and $\mathrm{L2G}(\cdot)$ represents the transformation process of the L2G block.

$\mathrm{L2G}(\cdot)$ takes the local features extracted by the CNN branch as input and reduces their dimensionality to match the input dimensionality of the ViT branch. Specifically, it first performs average pooling on the local features to obtain a $Batch\_size \times Channel\_num$ tensor $F_{Mean}$, where $Batch\_size$ represents the number of samples in the current batch and $Channel\_num$ represents the number of feature dimensions in the current CNN layer.
Therefore, $F_L^k$ in the local to global block first goes through the operation of $\mathrm{Mean}(\cdot)$, as shown in the following Equation (\ref{eq11}).
\begin{equation}\label{eq11}
F_{Mean} = \mathrm{Mean}(F_L^k),
\end{equation}
where, the input of the $\mathrm{Mean}(\cdot)$ operation is $F_L^k$, while the output is $F_{Mean}$.

$\mathrm{Mean}(\cdot)$ performs average pooling on the local feature tensor $F_L^k$ along the last two dimensions, resulting in a tensor $F_{Mean}^k$ with a shape of $32 \times 64$, where $32$ represents the batch size and $64$ represents the dimensionality of the local features extracted by the CNN.  The $\mathrm{Mean}( \cdot )$ operation is shown in the following Equation (\ref{eq12}).

\begin{equation}\label{eq12}
\mathrm{Mean}(F_L^k) = \frac{1}{{h \times w}}{\sum\limits_{m = 1}^h {\sum\limits_{n = 1}^w {F^k_L{}_{m,n}}}},k \in \left\{ {1,2,3} \right\},
\end{equation}
where $h$ represents the height of the feature map, and $w$ represents the width of the feature map.

Then, to facilitate fusion with global features for better performance, $\mathrm{L2G}(\cdot)$ needs to map the tensor $F_{Mean}$ to a space with the same dimensionality as the global features. Specifically, $\mathrm{L2G}(\cdot)$ applies a linear mapping to $F_{Mean}$ using the $\mathrm{Linear}(\cdot)$ operation, resulting in a tensor $F_{Linear}^k$ with a shape of $32 \times 768$, which has the same dimensionality as the global features. The $\mathrm{Linear}( \cdot )$ operation is shown in the following Equation (\ref{eq13}) and Equation (\ref{eq14}).

\begin{equation}\label{eq13}
F_{Linear} = \mathrm{Linear}(F_{Mean})
\end{equation}
\begin{equation}\label{eq14}
\mathrm{Linear}(F_{Mean}) = {W^T}F_{Mean} + b,
\end{equation}
where, the input of the $\mathrm{Linear}( \cdot )$ operation is $F_{Mean}$, while the output is $F_{Linear}$, ${W^T}$ is the linear transformation matrix, and $b$ is the deviation value.

Finally, to preserve the original global features, $\mathrm{L2G}(\cdot)$ fuses the global feature $F_G^k$ of stage $k$ with $F_{Linear}$ to obtain the input of the Transformer branch for stage $k+1$, namely the global feature $F_G^{k + 1}$ of stage $k+1$. Therefore, to better combine with $F_G^k$ of stage $k$ global feature matrix, $\mathrm{L2G}(\cdot)$ first repeats the tensor $F_{Linear}$ along the specified dimension to expand its shape. Specifically, $\mathrm{Expand}(\cdot)$ is used to copy the tensor $F_{Linear}$ along the second dimension so that it matches the global feature $F_G^k$ of stage $k$ in this dimension. Then, the Concat function is used to concatenate the expanded tensor $F_{Linear}$ with the global feature $F_G^k$ of stage $k$ along the second dimension. The fusion operation is shown in the following Equation (\ref{eq15}).

\begin{equation}\label{eq15}
F_G^{k + 1} = \mathrm{Expand}(F_{Linear})\parallel F_G^k,k \in \left\{ {1,2,3} \right\},
\end{equation}
where $\parallel$ represents the concatenate operation of the Concat function, $\mathrm{Expand}(\cdot)$ represents the expansion operation.

\subsubsection{G2L Block}

The G2L block is responsible for transforming the input ViT feature maps into CNN feature maps. This block consists of Squeeze, Unsqueeze, Expand, Conv $1\times1$, and Concat. The processing of the G2L block is shown in the following Equation (\ref{eq16}). 

\begin{equation}\label{eq16}
F_L^{k + 1} = \mathrm{G2L}(F_G^k), k \in \left\{ {1,2,3} \right\},
\end{equation}
where $F_L^{k + 1}$ represents the local feature at stage $k+1$ and serves as the output to the G2L block, $F_G^k$ represents the global feature at stage $k$ and serves as the input of the G2L block, and $\mathrm{G2L}(\cdot)$ represents the transformation process of the G2L block. Specifically, $\mathrm{G2L}(\cdot)$ is a feature extractor that operates from global to local and extracts local information from the feature map of ViT. During this process, $\mathrm{G2L}(\cdot)$ first uses the $\mathrm{Squeeze}(\cdot)$ and $\mathrm{Unsqueeze}(\cdot)$ functions to manipulate the dimensions of the ViT feature map. 

$\mathrm{G2L}(\cdot)$ first compresses one dimension of the tensor $F_G^k$ along its second dimension using the $\mathrm{Squeeze}(\cdot)$ function, and extracts the feature vector of the first channel, resulting in a tensor $F_{Squeeze}$ with a shape of $32 \times 768$, where $32$ represents the number of samples and $768$ represents the number of channels, as shown in the following equation Equation (\ref{eq17}).
\begin{equation}\label{eq17}
F_{Squeeze} = \mathrm{Squeeze}(F_G^k),
\end{equation}
where, the input of the $\mathrm{Squeeze}(\cdot)$ operation is $F_G^k$, while the output is $F_{Squeeze}$.

In order to better fuse with the local features extracted by the CNN branch and obtain better feature representations, $\mathrm{G2L}(\cdot)$ further transforms the shape of the tensor $F_{Squeeze}$ from $32 \times 768$ to a four-dimensional tensor with shape $32 \times 768 \times 8 \times 8$. Specifically, $\mathrm{G2L}(\cdot)$ first uses two $\mathrm{Unsqueeze}(\cdot)$ operations to add two dimensions to the $F_{Squeeze}$ tensor, resulting in a three-dimensional tensor with shape $32 \times 768 \times 1 \times 1$. Then, the $\mathrm{Expand}(\cdot)$ operation is used to expand this tensor along the last two dimensions to a four-dimensional tensor with shape $32 \times 768 \times 8 \times 8$, as shown in the following equation Equation (\ref{eq18}).

\begin{equation}\label{eq18}
\resizebox{.8\hsize}{!}{$F_{Expand} = \mathrm{Expand}(\mathrm{Unsqueeze}(\mathrm{Unsqueeze}(F_{Squeeze})))$},
\end{equation}

Where, the input to the equation is $F_{Squeeze}$ and the output is $F_{Expand}$, $\mathrm{Unsqueeze}(\cdot)$ adds dimensions to the tensor, while $\mathrm{Expand}(\cdot)$ performs an expanding operation.

In order to enhance the representational power of global features while preserving local feature information to improve the model's performance, $\mathrm{G2L}(\cdot)$ fuses the local features extracted from the CNN branch with global features. Specifically, $\mathrm{G2L}(\cdot)$ concatenates the local feature $F_L^{k}$ extracted from the CNN branch with the globally transformed feature $F_{Expand}$ along the channel dimension, resulting in a tensor $F_{Concat}$ with a shape of $32 \times (64+768) \times 8 \times 8$, where 32 represents the batch size, 64 represents the channel number of $F_L^{k}$, 768 represents the channel number of $F_{Expand}$, and $8 \times 8$ represents the height and width of the tensor, as shown in the following equation Equation (\ref{eq19}).
\begin{equation}\label{eq19}
F_{Concat} = F_{Expand}\parallel F_L^{k},k \in \left\{ {1,2,3} \right\},
\end{equation}
where $\parallel$ represents the concatenate operation of the Concat function. Then, the $1\times1$ convolutional layer $\mathrm{Conv2d}(\cdot)$ is applied to $F_{Concat}$ to perform convolution and obtain a tensor $F_L^{k+1}$ with a shape of $32 \times 64 \times 8 \times 8$, which achieves the fusion of local and global features. The final feature $F_L^{k+1}$ is used as the input of the CNN branch for stage $k+1$, as shown in the following equation Equation (\ref{eq20}).
\begin{equation}\label{eq20}
F_L^{k+1} = \mathrm{Conv2d}(F_{Concat}),k \in \left\{ {1,2,3} \right\},
\end{equation}
where, the input of the $\mathrm{Conv2d}(\cdot)$ operation is $F_{Concat}$, while the output is $F_L^{k+1}$.

\section{EXPERIMENTS}
In this section, we first introduce two widely used datasets. Then, the experiment settings are described. Finally, the results of the datasets are reported and discussed.

\subsection{Datasets and settings}

We trained and tested the CIT-EmotionNet model using Tesla V100-SXM2-32GB GPU, and implemented it using the PyTorch framework. The training was conducted using Adam optimizer, and the learning rate was set to 1e-5. The batch size was set to 32, and the dropout rate was set to 0.2. The number of classes to classify for the SEED dataset was 3, while for the SEED-IV dataset, it was 4. The samples were randomly shuffled, and the data were divided into training and testing sets with a ratio of 6:4. The cross-entropy loss function was used. 

\subsection{Experimental Results and Analysis}
In order to validate the effectiveness of our proposed model, we compared it with several baseline methods on the SEED datasets and SEED IV datasets, and briefly introduced each method as follows.

Table~\ref{The performance on the SEED and SEED IV datasets} reports the average accuracy(ACC) and standard deviation(STD) of these methods and the proposed CIT-EmotionNet model for EEG emotion recognition. The SST-EmotionNet utilized a 3D CNN-based method and employed spatial-spectral-temporal features, achieving an ACC of 96.02\% and 84.92\% on the SEED dataset and SEED IV dataset, respectively. However, the SST-EmotionNet did not consider global features. EeT used a Transformer-based method and thus performed better than CNN-based methods on the large SEED dataset, achieving an ACC of 96.28\%. However, EeT ignored local features and performed worse on the smaller SEED IV dataset, achieving an ACC of 83.27\%. MDGCN-SRCNN is a combined method based on GCN and CNN that considers the fusion of different features, achieving an ACC of 95.08\% and 85.22\% on the SEED dataset and SEED IV dataset, respectively. In general, CIT-EmotionNet integrates local and global features well, enabling it to capture valuable features from EEG signals for emotion recognition comprehensively. Compared to the baseline model, CIT-EmotionNet further improves its accuracy. It achieved an ACC of 98.57\% and an STD of 1.38\% on the SEED dataset and an ACC of 92.09\% and an STD of 5.33\% on the SEED IV dataset, outperforming the current state-of-the-art methods.

\begin{table*}[ht]
  \caption{Performance comparison between the baseline methods and the proposed CIT-EmotionNet on the SEED and SEED-IV datasets.\label{The performance on the SEED and SEED IV datasets}}
  \centering
  \begin{tabular}{cccccc}
    \toprule

\multirow{2}*{Method} & \multirow{2}*{Year}& \multicolumn{2}{c}{SEED} & \multicolumn{2}{c}{SEED-IV} \\

\cmidrule(lr){3-4}\cmidrule(lr){5-6}

&   & \textbf{ACC (\%)} $\uparrow $ & \textbf{STD (\%)} $\downarrow$ & \textbf{ACC (\%)} $\uparrow $  & \textbf{STD (\%) $\downarrow$} \\\midrule

SVM\cite{zheng2015investigating}    & 2015    & 83.99 & 9.72      & 56.61     & 20.05     \\
DGCNN\cite{song2018eeg}        & 2018      & 90.40 & 8.49          & 69.88    & 16.29          \\
BiHDM\cite{li2020novel}     & 2019        & 93.12 & 6.06          & 74.35    & 14.09           \\
RGNN\cite{zhong2020eeg}         & 2020      & 94.24 & 5.95          &  79.37    & 10.54           \\
SST-EmotionNet\cite{jia2020sst}  & 2020    & 96.02  & 2.17          &  84.92    & 6.66           \\
3D CNN\&PST\cite{liu2021positional}  & 2021    & 95.76  & 4.98          &  82.73   & 8.96           \\
EeT\cite{liu2022spatial}  & 2021    & 96.28  & 4.39          &  83.27 & 8.37           \\
JDAT\cite{wang2021jdat}  & 2021    & 97.30  & 1.74          &  - & -           \\
MD-AGCN\cite{li2021multi}     & 2021       & 94.81 & 4.52          & 87.63 & 5.77           \\
4D-aNN\cite{xiao20224d}       & 2022      & 96.25 & 1.86          & 86.77 & 7.29           \\
MDGCN-SRCNN\cite{bao2022linking}  & 2022    & 95.08  & 6.12          &  85.52 & 11.58          \\
\hline
  \textbf{CIT-EmotionNet} & \textbf{2023}  & \textbf{98.57}   &   \textbf{1.38}    & \textbf{92.09}     &   \textbf{5.33}    \\
    \bottomrule
  \end{tabular}

\end{table*}

\subsection{Ablation Experiments}

In order to validate the effects of different components in our model on the EEG emotion recognition tasks, we performed ablation experiments on both the SEED and SEED IV datasets. The ablation experiments were conducted in two types. The first type involved the ablation of each major component in CIT-EmotionNet, aiming to validate the effectiveness of the fusion and interaction of local and global features. The second type involved ablation experiments on each component in the CIT module, aiming to validate the effectiveness of each component.

\subsubsection{Ablation experiments on the major components of CIT-EmotionNet}

CIT-EmotionNet consists of three main components: the CNN branch, the Transformer branch, and the CIT module. In order to validate the effectiveness of the fusion and interaction of local and global features, we conducted ablation experiments on the major components of CIT-EmotionNet. Table~\ref{Ablation experiments on the major components of CIT-EmotionNet} illustrates the impact of different components of CIT-EmotionNet on EEG emotion recognition tasks. "With only CNN branch" indicates the use of only the CNN branch, which means using ResNet alone for the EEG emotion recognition task. "With only Transformer branch" indicates the use of only the Transformer branch, which means using ViT alone for the EEG emotion recognition task. "Baseline" refers to the removal of all CIT modules, and only a simple concatenation fusion is performed on the output of the feature by the CNN branch, and the Transformer branch.

"With only CNN branch" achieves an ACC of 74.36\% and an STD of 8.72\% on SEED datasets, while achieving an ACC of 50.93\% and an STD of 18.56\% on SEED IV datasets. "With only Transformer branch" achieves an ACC of 91.77\% and an STD of 3.98\% on SEED datasets, while achieving an ACC of 82.87\% and an STD of 7.18\% on SEED IV datasets. This indicates that Transformer performs better than CNN on the EEG emotion recognition task. "Baseline" achieved an accuracy of 93.39\% and a standard deviation of 3.66\% on the SEED dataset, while on the SEED IV dataset, the accuracy and standard deviation were 87.81\% and 6.63\%, respectively. These results indicate that the combination of local and global features contributes to improving the recognition performance of the model. 

\begin{table*}[ht]
\caption{Ablation experiments on the major components of CIT-EmotionNet.\label{Ablation experiments on the major components of CIT-EmotionNet}}
  \centering
  \begin{tabular}{ccccc}
    \toprule

\multirow{2}*{Method} & \multicolumn{2}{c}{SEED} & \multicolumn{2}{c}{SEED-IV} \\

\cmidrule(lr){2-3}\cmidrule(lr){4-5}

&   \textbf{ACC (\%)} $\uparrow $ & \textbf{STD (\%) } $\downarrow$ & \textbf{ACC (\%)}  $\uparrow $ & \textbf{STD (\%)} $\downarrow$\\

\midrule

With only CNN branch   & 74.36 & 8.72      & 50.93 & 18.56     \\
With only Transformer branch      & 91.77 & 3.98           & 82.87 & 7.18           \\
Baseline     & 93.39 & 3.66          & 87.81 & 6.63           \\

  \textbf{CIT-EmotionNet}  & \textbf{98.57}   &   \textbf{1.38}    & \textbf{92.09}     &   \textbf{5.33}    \\
\bottomrule
\end{tabular}
\end{table*}

\subsubsection{Ablation experiments on the major components of CIT module}

To validate the effectiveness of the L2G and G2L blocks in the CIT module, ablation experiments were conducted on both blocks. Additionally, ablation experiments were performed to investigate the impact of the Transformer feature map and CNN feature map on the L2G and G2L blocks, respectively. Table~\ref{Ablation experiments on the major components of CIT module} presents the effects of the major components of the CIT module on EEG emotion recognition tasks. "With only L2G block" refers to the integration of Local features into Global features, along with the fusion of Transformer Feature Maps in each stage. "With only L2G block and w/o TFM" means that only the interaction between local and global features was added on top of concatenating the output of the feature by the CNN and Transformer branches, without fusing the Transformer feature map (TFM) at each stage. "With only G2L block" refers to the integration of Global features into Local features, along with the fusion of CNN Feature Maps in each stage. "With only G2L block and w/o CFM" means that only the interaction between global and local features was added on top of concatenating the output of the feature by the CNN and Transformer branches, without fusing the CNN feature map (CFM) at each stage. 

"With only L2G block", the ACC was 96.92\% and 90.37\% on the SEED dataset and SEED IV dataset, respectively, indicating that using only L2G block cannot achieve satisfactory recognition performance. On the other hand, "With only L2G block and w/o TFM", the ACC was 78.83\% and 58.18\% on the SEED dataset and SEED IV dataset, respectively, suggesting that incorporating TFM in L2G block can improve the accuracy of the model. "With only G2L block", the ACC on the SEED dataset and SEED IV dataset were 97.53\% and 91.29\%, respectively, indicating that using only the G2L block cannot achieve satisfactory recognition performance. "With only G2L block and w/o CFM", the ACC on the SEED dataset and SEED IV dataset were 96.65\% and 89.39\%, respectively, suggesting that incorporating CFM in the G2L block can improve the accuracy of the model.

\begin{table*}[ht]
\caption{Ablation experiments on the major components of CIT module.\label{Ablation experiments on the major components of CIT module}}
\centering
  \begin{tabular}{ccccc}
    \toprule

\multirow{2}*{Method} & \multicolumn{2}{c}{SEED} & \multicolumn{2}{c}{SEED-IV} \\

\cmidrule(lr){2-3}\cmidrule(lr){4-5}

&   \textbf{ACC (\%)} $\uparrow $ & \textbf{STD (\%)} $\downarrow$ & \textbf{ACC (\%)} $\uparrow $ & \textbf{STD (\%)} $\downarrow$\\

\midrule

With only L2G block      & 96.92 & 2.08           & 90.37 & 6.13           \\
With only L2G block and w/o TFM     & 78.83 & 7.49           & 58.18 & 11.35           \\
With only G2L block      & 97.53 & 1.81          & 91.29\ & 5.62 
\\
With only G2L block and w/o CFM      & 96.65 & 3.04          & 89.39\ & 6.29          \\

  \textbf{CIT-EmotionNet}  & \textbf{98.57}   &   \textbf{1.38}    & \textbf{92.09}     &   \textbf{5.33}    \\
\bottomrule
\end{tabular}
\end{table*}

\section{CONCLUSION}

In this paper, we propose a novel CNN interactive Transformer Network for EEG Emotion Recognition, referred to as CIT-EmotionNet, which effectively integrates the local and global features of EEG signals and uses a parallel approach of CNN and Transformer for EEG emotion recognition tasks. Firstly, we transformed the PSD features of EEG signals into spatial-frequency representations, which were used as the input of the proposed model. Then, to extract both local and global features from EEG signals simultaneously, we designed a parallel structure of CNN and Transformer, which were unified in the same framework. Finally, we developed the CIT module to facilitate the interaction and fusion of local and global features, thereby enhancing the ability of the model to extract local and global characteristics from EEG spatial-frequency representations. The CIT module was utilized in three stages of CNN and Transformer. The proposed CIT-EmotionNet achieved average recognition accuracy of 98.57\% and 92.09\% on the SEED and SEED-IV datasets, respectively, outperforming the state-of-the-art methods. To verify the effects of different components in CIT-EmotionNet on the EEG emotion recognition task, we conducted ablation experiments on the SEED and SEED-IV datasets. The experimental results demonstrated that the CIT module was beneficial to the fusion and interaction of local and global features, leading to an improvement in the recognition performance.

{\small
\bibliographystyle{ieee_fullname}
\bibliography{main}

\begin{thebibliography}{10}\itemsep=-1pt

\bibitem{atkinson2016improving}
John Atkinson and Daniel Campos.
\newblock Improving bci-based emotion recognition by combining eeg feature
  selection and kernel classifiers.
\newblock {\em Expert Systems with Applications}, 47:35--41, 2016.

\bibitem{bao2022linking}
Guangcheng Bao, Kai Yang, Li Tong, Jun Shu, Rongkai Zhang, Linyuan Wang, Bin
  Yan, and Ying Zeng.
\newblock Linking multi-layer dynamical gcn with style-based recalibration cnn
  for eeg-based emotion recognition.
\newblock {\em Frontiers in Neurorobotics}, 16, 2022.

\bibitem{chen2019accurate}
JX Chen, PW Zhang, ZJ Mao, YF Huang, DM Jiang, and YN Zhang.
\newblock Accurate eeg-based emotion recognition on combined features using
  deep convolutional neural networks.
\newblock {\em IEEE Access}, 7:44317--44328, 2019.

\bibitem{cimtay2020cross}
Yucel Cimtay, Erhan Ekmekcioglu, and Seyma Caglar-Ozhan.
\newblock Cross-subject multimodal emotion recognition based on hybrid fusion.
\newblock {\em IEEE Access}, 8:168865--168878, 2020.

\bibitem{doma2020comparative}
Vikrant Doma and Matin Pirouz.
\newblock A comparative analysis of machine learning methods for emotion
  recognition using eeg and peripheral physiological signals.
\newblock {\em Journal of Big Data}, 7(1):1--21, 2020.

\bibitem{jia2020sst}
Ziyu Jia, Youfang Lin, Xiyang Cai, Haobin Chen, Haijun Gou, and Jing Wang.
\newblock Sst-emotionnet: Spatial-spectral-temporal based attention 3d dense
  network for eeg emotion recognition.
\newblock In {\em Proceedings of the 28th ACM international conference on
  multimedia}, pages 2909--2917, 2020.

\bibitem{jia2021hetemotionnet}
Ziyu Jia, Youfang Lin, Jing Wang, Zhiyang Feng, Xiangheng Xie, and Caijie Chen.
\newblock Hetemotionnet: two-stream heterogeneous graph recurrent neural
  network for multi-modal emotion recognition.
\newblock In {\em Proceedings of the 29th ACM International Conference on
  Multimedia}, pages 1047--1056, 2021.

\bibitem{kwon2018electroencephalography}
Yea-Hoon Kwon, Sae-Byuk Shin, and Shin-Dug Kim.
\newblock Electroencephalography based fusion two-dimensional (2d)-convolution
  neural networks (cnn) model for emotion recognition system.
\newblock {\em Sensors}, 18(5):1383, 2018.

\bibitem{li2021multi}
Rui Li, Yiting Wang, and Bao-Liang Lu.
\newblock A multi-domain adaptive graph convolutional network for eeg-based
  emotion recognition.
\newblock In {\em Proceedings of the 29th ACM International Conference on
  Multimedia}, pages 5565--5573, 2021.

\bibitem{li2022eeg}
Xiang Li, Yazhou Zhang, Prayag Tiwari, Dawei Song, Bin Hu, Meihong Yang,
  Zhigang Zhao, Neeraj Kumar, and Pekka Marttinen.
\newblock Eeg based emotion recognition: A tutorial and review.
\newblock {\em ACM Computing Surveys}, 55(4):1--57, 2022.

\bibitem{li2020novel}
Yang Li, Lei Wang, Wenming Zheng, Yuan Zong, Lei Qi, Zhen Cui, Tong Zhang, and
  Tengfei Song.
\newblock A novel bi-hemispheric discrepancy model for eeg emotion recognition.
\newblock {\em IEEE Transactions on Cognitive and Developmental Systems},
  13(2):354--367, 2020.

\bibitem{liu2022spatial}
Jiyao Liu, Hao Wu, Li Zhang, and Yanxi Zhao.
\newblock Spatial-temporal transformers for eeg emotion recognition.
\newblock In {\em 2022 The 6th International Conference on Advances in
  Artificial Intelligence}, pages 116--120, 2022.

\bibitem{liu2021positional}
Jiyao Liu, Yanxi Zhao, Hao Wu, and Dongmei Jiang.
\newblock Positional-spectral-temporal attention in 3d convolutional neural
  networks for eeg emotion recognition.
\newblock In {\em 2021 Asia-Pacific Signal and Information Processing
  Association Annual Summit and Conference (APSIPA ASC)}, pages 305--312. IEEE,
  2021.

\bibitem{salama2018eeg}
Elham~S Salama, Reda~A El-Khoribi, Mahmoud~E Shoman, and Mohamed A~Wahby
  Shalaby.
\newblock Eeg-based emotion recognition using 3d convolutional neural networks.
\newblock {\em International Journal of Advanced Computer Science and
  Applications}, 9(8), 2018.

\bibitem{shen2023triplet}
Fei Shen, Xiaoyu Du, Liyan Zhang, and Jinhui Tang.
\newblock Triplet contrastive learning for unsupervised vehicle
  re-identification.
\newblock {\em arXiv preprint arXiv:2301.09498}, 2023.

\bibitem{git}
Fei Shen, Yi Xie, Jianqing Zhu, Xiaobin Zhu, and Huanqiang Zeng.
\newblock Git: Graph interactive transformer for vehicle re-identification.
\newblock {\em IEEE Transactions on Image Processing}, 2023.

\bibitem{hpgn}
Fei Shen, Jianqing Zhu, Xiaobin Zhu, Yi Xie, and Jingchang Huang.
\newblock Exploring spatial significance via hybrid pyramidal graph network for
  vehicle re-identification.
\newblock {\em IEEE Transactions on Intelligent Transportation Systems}, 2021.

\bibitem{song2018eeg}
Tengfei Song, Wenming Zheng, Peng Song, and Zhen Cui.
\newblock Eeg emotion recognition using dynamical graph convolutional neural
  networks.
\newblock {\em IEEE Transactions on Affective Computing}, 11(3):532--541, 2018.

\bibitem{tan2020fusionsense}
Clarence Tan, Gerardo Ceballos, Nikola Kasabov, and Narayan
  Puthanmadam~Subramaniyam.
\newblock Fusionsense: Emotion classification using feature fusion of
  multimodal data and deep learning in a brain-inspired spiking neural network.
\newblock {\em Sensors}, 20(18):5328, 2020.

\bibitem{wang2021jdat}
Zeyu Wang, Ziqun Zhou, Haibin Shen, Qi Xu, and Kejie Huang.
\newblock Jdat: Joint-dimension-aware transformer with strong flexibility for
  eeg emotion recognition.
\newblock 2021.

\bibitem{xiao20224d}
Guowen Xiao, Meng Shi, Mengwen Ye, Bowen Xu, Zhendi Chen, and Quansheng Ren.
\newblock 4d attention-based neural network for eeg emotion recognition.
\newblock {\em Cognitive Neurodynamics}, pages 1--14, 2022.

\bibitem{xing2019sae+}
Xiaofen Xing, Zhenqi Li, Tianyuan Xu, Lin Shu, Bin Hu, and Xiangmin Xu.
\newblock Sae+ lstm: A new framework for emotion recognition from multi-channel
  eeg.
\newblock {\em Frontiers in neurorobotics}, 13:37, 2019.

\bibitem{yang2019multi}
Heekyung Yang, Jongdae Han, and Kyungha Min.
\newblock A multi-column cnn model for emotion recognition from eeg signals.
\newblock {\em Sensors}, 19(21):4736, 2019.

\bibitem{zheng2015investigating}
Wei-Long Zheng and Bao-Liang Lu.
\newblock Investigating critical frequency bands and channels for eeg-based
  emotion recognition with deep neural networks.
\newblock {\em IEEE Transactions on autonomous mental development},
  7(3):162--175, 2015.

\bibitem{zhong2020eeg}
Peixiang Zhong, Di Wang, and Chunyan Miao.
\newblock Eeg-based emotion recognition using regularized graph neural
  networks.
\newblock {\em IEEE Transactions on Affective Computing}, 13(3):1290--1301,
  2020.

\end{thebibliography}
}

\end{document}